\documentclass[journal]{IEEEtran}
\usepackage{cite}
\usepackage{amsmath,amssymb,amsfonts}
\usepackage{nicefrac}
\usepackage{siunitx}
\usepackage{booktabs}
\usepackage{nicematrix}
\usepackage{multirow}
\usepackage{algorithmic}
\usepackage{graphicx}
\usepackage{textcomp}
\usepackage{xcolor}
\usepackage[colorlinks=true,bookmarks=false,citecolor=blue,urlcolor=black,linkcolor=red]{hyperref} %
\usepackage{pgfplots}
\usepackage[nolist]{acronym} 
\pgfplotsset{compat=1.18}
\pgfplotsset{
/pgfplots/short line/.style={
    legend image code/.code={
        \draw[mark repeat=2,mark phase=2,#1] 
            plot coordinates {(0cm,0cm) (0.1275cm,0cm) (0.245cm,0cm)};
    },
},
}
\pgfdeclareplotmark{hexagon}
{
\newdimen\R
\R=2.25pt
\draw (0:\R) \foreach \x in {60,120,...,300} {
        -- (\x:\R)
    }-- cycle (90:\R);
}
\usetikzlibrary{arrows,shapes,graphs,graphs.standard,quotes,arrows.meta,decorations.markings,positioning,fit,positioning,hobby,decorations.text,decorations.pathreplacing}

\usepackage{transparent}
\usepackage{soul}
\usepackage{bm}

\definecolor{mittelblau}{RGB}{0, 126, 198}
\definecolor{rot}{RGB}{238, 28 35}
\definecolor{apfelgruen}{RGB}{140, 198, 62}
\definecolor{gelb}{RGB}{255, 229, 0}
\definecolor{orange}{RGB}{244, 111, 33}
\definecolor{pink}{RGB}{237, 0, 140}
\definecolor{lila}{RGB}{128, 10, 145}
\definecolor{hellgrau}{RGB}{224, 224, 224}
\definecolor{mittelgrau}{RGB}{128, 128, 128}
\definecolor{dunkelgrau}{RGB}{80,80,80}
\definecolor{anthrazit}{RGB}{19, 31, 31}
\definecolor{darkgreen}{RGB}{34,139,34}

\newcommand{\dv}{d_{\mathsf{v}}}
\newcommand{\dc}{d_{\mathsf{c}}}

\begin{acronym}
    \acro{AED}{automorphism ensemble decoding}
    \acro{ASIC}{application-specific integrated circuit}
    \acro{AWGN}[AWGN]{additive white Gaussian noise}
    \acro{BCH}{Bose--Chaudhuri--Hocquenghem}
    \acro{BEC}[BEC]{binary erasure channel}
    \acro{BER}{bit error rate}
    \acro{BP}{belief propagation}
    \acro{BSC}[BSC]{binary symmetric channel}
    \acro{CN}{check node}
    \acro{CRC}{cyclic redundancy check}
    \acro{CA}{\ac{CRC}-aided}
    \acro{DVB}{Digital Video Broadcasting -- Satellite}
    \acro{DVB-S2}{Digital Video Broadcasting -- Satellite 2nd Generation}
    \acro{DSP}[DSP]{digital signal processing}
    \acro{eMBB}{enhanced mobile broadband}
    \acro{EXIT}{extrinsic information transfer}
    \acro{FER}{frame error rate}
    \acro{HARQ}{hybrid automatic repeat request}
    \acro{HRLLC}{hyper reliable low latency communications}
    \acro{IoT}{internet of things}
    \acro{KPI}{key performance indicator}
    \acro{LBP}{Layered Belief Propagation}
    \acro{LDPC}{low-density parity-check}
    \acro{LTE}[LTE]{Long Term Evolution}
    \acro{MAP}[MAP]{maximum a posteriori}
    \acro{ML}{maximum likelihood}
    \acro{mMTC}{massive machine type communication}
    \acro{mULC}{massive ultra-reliable low-latency communication}
    \acro{NMS}{normalized min-sum}
    \acro{OMS}{offset min-sum}
    \acro{PAS}{probabilistic amplitude shaping}
    \acro{PBRL}{protograph-based raptor-like}
    \acro{PCM}{parity check matrix}
    \acrodefplural{PCM}[PCMs]{parity check matrices}
    \acro{PFT}{polar factor tree}
    \acro{QC}{quasi-cyclic}
    \acro{RS}{Reed--Solomon}
    \acro{SC}{successive cancellation}
    \acro{SC-LDPC}{spatially-coupled low-density parity-check}
    \acro{SCL}{successive cancellation list}
    \acro{SCF}{successive cancellation flip}
    \acro{SED}{scheduling ensemble decoder}
    \acro{SNR}{signal-to-noise ratio}
    \acro{SSC}{simplified successive cancellation}
    \acro{SPA}{sum-product algorithm}
    \acro{ULBC}{ultra-reliable low-latency broadband communication}
    \acro{uMBB}{ubiquitous mobile broadband}
    \acro{VN}{variable node}
\end{acronym}
\begin{document}

\title{Towards a Unified Coding Scheme for 6G \\[-2.5em] {\large An Open6GHub White Paper} \\[1.5em]
    \thanks{
        The authors acknowledge the financial support by the Federal Ministry of
        Education and Research of Germany in the project “Open6GHub” (grant
        numbers: 16KISK004, 16KISK010, and 16KISK019).
    }
}

\author{Paul Bezner, Erdem Eray Cil, Jannis Clausius, Oliver Griebel, Tim Janz, Lucas Johannsen, Claus Kestel, \\ Felix Krieg,  Haizheng Li, Jonathan Mandelbaum, Sisi Miao, Marvin Rübenacke, Daniel Tandler, Andreas Zunker \\ 
Laurent Schmalen, Norbert Wehn, Stephan ten Brink 
\thanks{Paul Bezner, Jannis Clausius, Tim Janz, Felix Krieg, Marvin Rübenacke, Daniel Tandler, Andreas Zunker, and Stephan ten Brink are with the Institute of Telecommunications, University of Stuttgart, 70569 Stuttgart, Germany. Email: \texttt{\{lastname\}@inue.uni-stuttgart.de}.

Oliver Griebel, Lucas Johannsen, Claus Kestel, and Norbert Wehn are with the Microelectronic Systems Design Research Group, University of Kaiserslautern (RPTU), 67663 Kaiserslautern, Germany. Email: \texttt{\{firstname.lastname\}@rptu.de}. Lucas Johannsen is also with the Koblenz University of Applied Sciences, 56075 Koblenz, Germany. Email: \texttt{johannsen@hs-koblenz.de}.

Erdem Eray Cil, Haizheng Li, Jonathan Mandelbaum, Sisi Miao, and Laurent Schmalen are with the Karlsruhe Institute of Technology (KIT), Communications Engineering Lab (CEL), Hertzstr. 16, 76187 Karlsruhe, Germany. E-mail: \{\texttt{firstname.lastname\}@kit.edu}}}

\maketitle

\begin{abstract}
The growing demand for higher data rates necessitates continuous innovations in wireless communication systems, particularly with the emergence of 6G. Channel coding plays a crucial role in this evolution. In 5G systems, rate-adaptive raptor-like quasi-cyclic irregular low-density parity-check codes are used for the data link, while polar codes with successive cancellation list decoding handle short messages on the synchronization channel. However, to meet the stringent requirements of future 6G systems, a versatile and unified coding scheme should be developed---one that offers competitive error-correcting performance alongside low complexity encoding and decoding schemes that enable energy-efficient hardware implementations.

This white paper outlines the vision for such a unified coding scheme. We explore various 6G communication scenarios that pose new challenges to channel coding and provide a first analysis of potential solutions.
\end{abstract}

\begin{IEEEkeywords}
wireless communications, channel coding, 6G, LDPC codes, polar codes, unified coding scheme
\end{IEEEkeywords}

\acresetall

\section{Introduction}\label{sec:introduction}
The development of advanced channel coding schemes has been a key driver in the evolution of wireless communication standards. From simple convolutional codes in GSM (2G) to parallel concatenated (turbo) codes in UMTS (3G) and LTE (4G), and later \ac{LDPC} and polar codes in 5G, every generation of the carefully designed channel codes introduced excellent improvements in performance, flexibility, and efficiency to the communication system. With 6G on the horizon, a unified coding scheme must be developed to support the next technology breakthrough. With recent advances in coding theory, next-generation error-correcting schemes are expected to deliver good error correction performance and address a wide range of new communication challenges. These challenges are mutually dependent and interconnected, sometimes with conflicting objectives, thus necessitating careful design of the coding scheme. In particular, coding gain alone is no longer the main figure of merit for system performance. Instead, the selected coding scheme must enable flexible configuration to be able to meet the requirements of a wide range of use cases, such as minimizing overall latency for time-critical applications or optimizing energy efficiency for, e.g., battery-powered \ac{IoT} devices~\cite{Rowshan2024}.

In this white paper, we will summarize the promises and challenges of 6G, motivating the use of a unified channel coding scheme. In particular, we discuss different promising technologies that are key to get a unified coding scheme based on polar or \ac{LDPC} codes.

\subsection{Promises of 6G}
Compared to 5G, the introduction of 6G comes with exciting new applications. For example, 6G will expand the scope of use cases, e.g., augmented reality applications, massive IoT networks for smart homes or factories, real-time data transmission for autonomous driving, and many more.

From the physical layer perspective, the new scenarios include \acf{eMBB}, \acf{HRLLC}, \acf{mMTC}, \acf{uMBB}, \acf{ULBC}, and \acf{mULC} \cite[p. 8-10]{Rowshan2024}\cite{uzunidis2023vision}. This reflects the diverse range of 6G applications, ranging from ultra-high-speed communication to extreme device density and energy-efficient systems, necessitating a flexible standard.

Additionally, due to the significant growth of the number of \ac{mMTC} connections, energy efficiency has become one of the major metrics of the channel coding scheme. Ideally, the new mobile devices should be both energy-efficient and future-proof, with significantly extended battery life and support for the long-term evolution of the technology.

\subsection{New System KPIs}
First of all, 6G introduces stricter ``hard'' \acp{KPI} compared to 5G to realize the promises, including higher throughput, decreased latency, and improved energy efficiency. In \cite{Chafii23Twelve, lu20206g, david20186g}, the authors present \acp{KPI} that are relevant for 6G channel codes. Compared to 5G, we expect to have 10x increase in cost efficiency, extended coverage, sub-ms decoding latency, 10x increase in energy efficiency, and 10-year lifetime of the battery-driven devices.
Additionally, we expect a \ac{BER} of $10^{-12}$ to be beneficial for compatibility with the Ethernet standard. Hence, the future coding scheme should not possess an error floor above a \ac{BER} of $10^{-12}$.

To meet these \acp{KPI}, the future coding scheme must offer flexibility and adaptability across different scenarios. For example, in \ac{IoT} or \ac{mMTC} use cases, energy efficiency and extended device lifetime are key parameters, whereas virtual reality or real-time autonomous systems, high data rate, ultra-low latency, and high reliability are required.

\subsection{Unified Coding Scheme}
To meet the increased demand for flexibility, one approach discussed in~\cite{bits2023unifiedcode} proposes the development of ``one-size-fits-all" channel codes. These codes can be deployed across different use cases with varying requirements, enabling flexibility and adaptability with minimal complexity. 

The concept of unified coding for 6G proposes a single, adaptable code family capable of meeting diverse use case requirements through adjustable decoding parameters. 
This approach necessitates a joint code-decoder design for all scenarios, simplifying 6G standards despite the challenge of satisfying all \acp{KPI} with a single code.

When faced with the task of designing a coding system for an application with a multitude of different scenarios---not just for 6G---developing separate coding schemes specifically tailored to each of these scenarios might yield the best results in terms of ``raw performance'', e.g., coding gain.
However, when aspects beyond this hard metric, such as sustainability, are taken into account, having a single and flexible family of channel codes, i.e., a \emph{unified code}~\cite{bits2023unifiedcode}, shared for all applications is a promising and elegant solution in particular to reduce costs in development and maintenance.

This unified approach brings numerous advantages that extend beyond traditional performance metrics. Key benefits include:
\begin{enumerate}
	\item {Excellent Communication Performance}:

With the advances in decoding algorithms with outstanding performance-complexity trade-offs, designing one coding scheme that offers excellent error-correcting performance for a wide range of scenarios is feasible.

	\item {Focus on Versatility}:
	
	A chosen coding scheme can be optimized to handle various environments and use cases effectively. This versatility can be more beneficial than having multiple schemes that excel in specific scenarios but underperform in others.
	
	\item {Design for efficient hardware implementation}: %
	Energy efficiency, throughput, latency, cost etc. require dedicated hardware implementations as \acp{ASIC} in advanced semiconductor technology nodes.  However the progress in semiconductor technology can not keep pace with the 6G requirements. Hence, the code and corresponding encoding/decoding scheme must permit efficient hardware implementation.
	
	\item {Easier Implementation and Lower Development Costs}:
	
	Developers can focus on optimizing a single coding scheme instead of multiple ones. This leads to more robust and reliable products while reducing development costs.
	
	\item {Sustainability Beyond Energy Efficiency}:
	
	The inherent flexibility of a unified code enables it to perform well in areas not originally foreseen during design time. Thus, such a 
	future-proof coding scheme increases the longevity of the standard and standard-conform devices.
	
	\item {Encouraging Innovation}:
	
	A simplified standard makes it easier for small companies to implement and produce standard-conform hardware. This is beneficial for driving innovation.
\end{enumerate}

In summary, future 6G codes and decoders must provide users with the flexibility to adapt to their specific needs. A unified coding is an enabling, future-proof technology for such a scheme. 

\subsection{Outline of the White Paper}
This white paper is organized as follows. In Sec.~\ref{sec:preliminaries}, we provide the necessary background for channel coding and define the notations. We also briefly review the construction of the \ac{LDPC} and polar codes employed in the 5G standard, respectively. Sec.~\ref{sec:limitation} illustrates the shortcomings of the two existing 5G channel coding schemes, while Sec.~\ref{sec:code_requirement} provides a comprehensive summary of the requirements of the new coding scheme. Then, we conduct extensive research on the comparison of the two most promising solutions, namely, the refined \ac{LDPC} and polar codes, described in Sec.~\ref{sec:polar_candidate} and Sec.~\ref{sec:LDPC_candidate}, respectively. Sec.~\ref{sec:conclusion} concludes the white paper.

\section{Preliminaries}\label{sec:preliminaries}
Throughout this paper, $(N,K)$ denotes a block code with block length $N$ and dimension $K$. The code rate will be denoted by $R=K/N$. We will also use $O(\cdot)$ for Landau's big-$O$ notation. Among the many families of linear codes, we focus on two prevailing candidates, i.e., \ac{LDPC} codes and polar codes.

\subsection{LDPC Codes}
A binary $(N,K)$ \ac{LDPC} code~\cite{Gal63} is defined as the null space of its (full rank) sparse \ac{PCM} $\boldsymbol{H}\in \mathbb{F}_2^{N-K\times N}$. Here, sparsity means that the number of $1$s, i.e., weights, of all rows and columns of $\boldsymbol{H}$ is upper-bounded by some constant which does not depend on the block length $N$. 
Note that the designed \ac{PCM} of \ac{LDPC} code is often overcomplete, i.e., it possesses $M>N-K$ rows, which potentially improves \ac{BP} decoding performance.
For a $(\dv, \dc)$ \textit{regular} \ac{LDPC} code, all row weights, denoted as $\dc$, are the same and all column weights, denoted as $\dv$, are also identical. An \ac{LDPC} code is \textit{irregular} if it does not have this property. A \ac{QC} \ac{LDPC} code features a \ac{PCM} consisting of an array of circulant matrices. The size of these circulant matrices is called the \textit{lifting size}.

The standard decoding algorithm for \ac{LDPC} codes is the \ac{BP} decoding algorithm, often denoted \ac{SPA}, which is a message-passing algorithm performed on the \textit{Tanner graph} associated with the code. The Tanner graph is a bipartite graph with two classes of vertices. The first class of vertices is the \acp{VN}, each corresponding to a code bit and thus to a column of $\boldsymbol{H}$. The second class of vertices is the \acp{CN}, each corresponding to a parity check and thus to a row of $\boldsymbol{H}$. An edge exists between a \ac{VN} and a \ac{CN} if the corresponding entry of the \ac{PCM} is $1$.
Besides the \ac{SPA}, there exist various simplified decoding variants, such as min-sum decoding. 

\subsection{Polar Codes}
Another important family of codes is polar codes~\cite{arikan2009}.
They are based on a process called \textit{channel polarization}, whereby a binary-input memoryless transmission channel is polarized into $N = 2^n$ synthetic bit channels by the polar transform $\boldsymbol G_N = \left[\begin{smallmatrix} 1 & 0 \\ 1 & 1 \end{smallmatrix}\right]^{\otimes n}$.
Most of these channels are polarized into highly reliable (``noiseless'') or highly unreliable (``noisy'') channels.
The ratio of noiseless channels to the block length $N$ approaches the channel capacity as $N$ tends to infinity.
To construct a binary $(N,K)$ polar code, the most reliable $K$ bit channels are selected to transmit information.
The indices of these information bits form the information set $\mathcal{I}$.
The remaining $N-K$ channels are frozen to a predefined value (typically zero).
They form the frozen set $\mathcal{F}$.
The selection of $\mathcal{I}$ can be optimized, e.g., by density evolution~\cite{trifonov2012DEPolar}. 
The generator matrix $\boldsymbol{G}$ of a polar code is found by selecting the rows of the polar transformation matrix $\boldsymbol{G}_N$ corresponding to the information bits $\mathcal{I}$.

Polar codes natively support rate adaptation through the selection of information bits.
To avoid relying on a unique bit channel reliability sequence for each code length, nested sequences can be constructed as used in the 5G standard~\cite{3GPP38.212}. 
However, code construction for arbitrary block lengths $N~\neq~2^n$ requires length-matching techniques such as puncturing or shortening. 

The original decoding algorithm for polar codes is \ac{SC} decoding, which sequentially estimates the bits transmitted over the synthetic channels from index $0$ to $N-1$, using the knowledge of the frozen bit values. 
Asymptotically, \ac{SC} decoding is optimal. However, to improve the performance in the finite block length regime, there exist more powerful variants of the \ac{SC} algorithm, including \ac{SCL} decoding~\cite{talvardy2015}, \ac{SCF} decoding \cite{afisiadis2014scf} and \ac{AED}~\cite{geiselhart2021polaraed}.

\section{Limitation of the 5G Codes}\label{sec:limitation}
The 5G standard employs \ac{LDPC} codes~\cite{design_of_5g_richie} for the data link channel and polar codes~\cite{3GPP14}\cite{Bioglio5Gpolar} for the control channel. 
The 5G \ac{LDPC} code is built around two distinct base matrices, each optimized for different rates and block lengths. It features a raptor-like, quasi-cyclic, irregular structure and uses state variables, i.e., punctured message bits, to meet 5G’s performance targets. However, these design elements face challenges in meeting the more stringent requirements expected from a future 6G system. The use of state variables, while improving threshold performance, requires a high number of decoding iterations, resulting in reduced efficiency and increased latency. Additionally, the highly irregular structure hinders the development of carefully optimized hardware implementations.

For an \ac{LDPC} code to be a strong contender for 6G, it must be more check-node regular and eliminate the need for state variables. This will allow for more flexible and efficient decoding architectures across a wider range of scenarios.

Similarly, the current 5G polar code design has significant limitations that must be addressed to remain a viable candidate for 6G. While the nested reliability sequence is an elegant solution, its current maximum code length of $1024$---with possible repetitions---limits its use in a wider range of applications. Additionally, the decoding process is inflexible, as the information set and \ac{CRC} concatenation are specifically tailored for \ac{SCL} decoding, hindering the use of other decoding algorithms such as \ac{AED} \cite{geiselhart2021polaraed}. 
Another concern is the shallow error slope of \ac{CRC}-aided polar codes \cite{rowmerged}, which drastically increases the required \ac{SNR} for highly reliable communication.
Hence, designing a \ac{CRC}-precoded polar code that meets some stricter performance demands of 6G, e.g., for \ac{HRLLC}, is challenging.

To be considered a strong candidate for 6G, the polar code design must support larger code lengths and incorporate a more universal information set, capable of supporting low-latency decoding techniques such as \ac{AED}. Furthermore, the outer code must be optimized to meet the rigorous requirements of \ac{HRLLC}, ensuring the reliability and low-latency performance that will be critical for some applications in 6G systems.

\section{Requirements for the 6G Code}\label{sec:code_requirement}
In Sec.~\ref{sec:introduction}, we described the high-level \acp{KPI} for the 6G channel coding scheme. In this section, we dive into the detailed technical requirements from both coding theory and hardware implementation perspectives. Because the individual requirements conflict with each other, it is challenging to satisfy all of them simultaneously. Therefore, we aim to identify a balanced ``sweet spot'' that offers an acceptable trade-off between performance, complexity, and feasibility.

\subsection{Performance Perspective}

\subsubsection{Universality}
A unified code must be flexible and support a wide range of applications to be universally applicable. To this end, the code design must be simple and not overfit to specific scenarios.

\subsubsection{Rate Adaptability}
 The wide range of 6G applications necessitates a channel error correction mechanism capable of operating across varying channel conditions. Therefore, rate adaptability is essential. For 6G, bit-wise granularity should be maintained for smaller block lengths; however, a coarser granularity for larger block lengths is expected to be sufficient.

\subsubsection{Hyper Reliability Scenario}
For hyper-reliable communication scenarios, the code must support lower error rates than 5G, potentially operating at a  target-\ac{BER} of $10^{-12}$ to allow for compatibility with Ethernet, while maintaining manageable complexity and scalability. 

\subsubsection{Low-Latency}
Time-critical applications demand minimal latency.
Thus, we require decoding algorithms with low latency. Additionally, the number of retransmissions should be minimized.
In particular, for iterative decoding schemes, the error correction performance for a limited number of decoding iterations must be considered at code design time besides the asymptotic performance.  

\subsubsection{Block Length Requirements}
Certain applications require short block length codes due to smaller payloads or more stringent requirements on latency, e.g., \ac{IoT} or control command communications, respectively. To this end,  codes with good performance in the short block length regime are necessary. 

Other applications, such as video streaming, require long block-length codes that provide better error correction performance at higher rates to meet throughput requirements. 
Yet, for large block lengths, the efficient hardware implementation of the decoding algorithm must remain feasible.

\subsubsection{Good Performance under Quantization}
As practical decoders use quantized messages, the code and decoder design must maintain the good performance with realistic bit-widths, e.g., 4-5 bits.
For instance, the target bit-widths should also be considered within the code design, e.g., by employing a density evolution that tracks quantized messages. The fact that the decoding algorithm can also correct quantization errors should be taken into account.

\subsubsection{Diversity Enabling}
For short block length scenarios, particularly in \ac{HRLLC}, the code or decoder should support diversity-enabling techniques to enhance error resilience, even at the cost of increased decoding complexity. Hereby, diversity-enabling refers to the possibility of improving decoding using an ensemble of constituent decoders, which potentially yields a list of candidate codewords. An ensemble is a set of constituent decoders (for the same code) that work non-cooperatively and can be built, e.g., by using automorphisms of the code or different decoding schedules. 

In conclusion, a unified 6G coding scheme must balance a wide range of requirements, including rate and block length adaptability, low latency, reliability, and diversity-enabling techniques. This holistic approach to code design is essential for addressing the diverse needs of 6G applications and ensuring feasible hardware implementations. The following section explores these concepts in greater detail.
        
\subsection{Hardware Implementation Perspective}\label{sec:iv.b.implementation_perspective}

From a hardware implementation perspective, the most challenging metrics for a 6G coding scheme are significantly reduced latency, higher throughput, and enhanced energy efficiency (defined as the energy required to decode a single bit) compared to 5G standards. These performance parameters are intrinsically linked to the characteristics of the underlying hardware implementation. If the maximum allowed power consumption is restricted, for example, due to the thermal design limitations of the package, any increase in throughput requires a corresponding proportional improvement in energy efficiency \cite{herrmann_energy_2021, zhang_channel_2023}. Thus, throughput enhancements inherently depend on improving energy efficiency. Additionally, power density, which directly affects thermal management and cooling requirements, is closely correlated with energy efficiency. Minimizing chip area is equally crucial, as this directly impacts manufacturing costs and integration efficiency.
The efficient hardware realization of 6G coding schemes is strongly determined by the specific complexity and algorithmic properties of the encoding and decoding processes. Algorithmic complexity can be categorized into three core components: computation, data transfer, and storage. Particularly in advanced technology nodes, data transfers and storage significantly outweigh computational operations concerning energy efficiency and power consumption. Different coding coding schemes exhibit varying profiles of these complexities. For example, decoding \ac{LDPC} codes is dominated by data transfers and storage due to its iterative process, whereas polar decoding achieves a more balanced distribution between computation and data transfers.
The second dimension involves algorithmic properties such as locality, regularity, parallelism, dataflow, and controlflow. These properties critically affect hardware architecture efficiency and overall design complexity. Algorithms characterized by high locality, regularity, and parallelism facilitate more efficient and scalable hardware architectures, simplifying implementation and enhancing performance. Algorithms oriented towards dataflow dominance are typically favored in hardware implementation due to their simpler control and improved hardware utilization compared to control-flow-driven approaches, which tend to increase design complexity and reduce utilization efficiency. However, coding schemes differ notably in these aspects. \Ac{LDPC} decoding algorithms based on belief propagation inherently offer a high degree of parallelism and regularity at the node level. Exploiting this parallelism is crucial for increasing throughput and reducing latency. However, it involves considerable data transfers due to its iterative nature and lacking locality due to the pseudo-randomness of the Tanner graph. Conversely, polar codes exhibit high structural regularity but require sequential behavior in the overall decoding process. Their sequential decoding behavior limits achievable throughput and latency reductions. Advanced polar decoding like list decoding implies control-flow due to the necessary sorting that decreases energy efficiency and increases area. Moreover, flexibility requirements like the support of different block lengths, code rates, or decoding schemes yield control-flow that reduces energy efficiency and increases area. Furthermore, adding flexibility such as multiple block lengths, various code rates or different coding schemes can introduce additional control flow, further impacting energy efficiency and chip area negatively. In conclusion, to meet the demanding performance and efficiency goals of 6G, early-stage co-optimization of code selection, algorithm structure, and hardware considerations is indispensable.

\subsection{Contribution of Numerical Simulations}
While theoretical analysis of code design and hardware interactions provides fundamental insights, simulations are critical for validating novel approaches and demonstrating practical performance gains. This work uses simulated error rate curves to quantify improvements (e.g., in \ac{FER} or \ac{BER} performance) and to highlight areas for optimization, bridging the gap between theoretical potential and real-world applicability.

The selection of code parameters (mainly block length and rate) of interest for emerging standards remains unclear, as the choices depend on channel scenarios and application-specific constraints that are yet to be defined. While the full parameter space is large---spanning different block lengths, rates, and decoder configurations---we focus on a few examples to illustrate interesting trade-offs in this work: 
\begin{itemize}  
    \item Block lengths: $N = 256$ to show the effects of short block lengths, and $N = 65536$ to show scalability beyond the code lengths typically considered in current wireless standards. 
    \item Rates: $R = \nicefrac{1}{2}$ and $R = \nicefrac{8}{9}$ to cover a wide range of reasonable rates.
\end{itemize}  
To contextualize the performance of practical codes, we compare the results to the meta-converse bound~\cite{PPV,SaddleApprox}. 
While coding schemes cannot achieve this (converse) bound, proximity to it quantifies the effectiveness of code-decoder pairs.  
To assess the complexity-performance trade-offs across code families, we evaluate several decoding algorithms for both \ac{LDPC} and polar codes, some under standardized configurations (\ac{DVB-S2} and 5G).
As shown in Fig.~\ref{fig:short_codes}, this analysis focuses on rate $\nicefrac{1}{2}$ codes with $N = 256$. 
For the \ac{LDPC} code, decoding follows the \ac{LBP} algorithm with $8$ iterations. In contrast, polar codes benefit from a wider range of decoders: We compare the low-complexity \ac{SC} baseline with two advanced variants---\ac{SCL} decoding (with list size $8$) and \ac{SC}-based \ac{AED} (using a different code design than 5G to allow for a richer automorphism group). The \ac{AED} decoder has an ensemble size of $8$, leading to a reasonably comparable algorithmic complexity between the \ac{LBP}, \ac{SCL}, and \ac{AED} decoders. We expect that real-world applications may also be based on complexities of a similar order of magnitude. 

Fig.~\ref{fig:long_medium_rate} and Fig.~\ref{fig:long_high_rate} focus on longer codes ($N=65536$). Fig.~\ref{fig:long_medium_rate} has a code rate of $\nicefrac{1}{2}$ and Fig.~\ref{fig:long_high_rate} has a rate of $\nicefrac{8}{9}$, relevant for scenario with very high throughput. In contrast to the short length figure, these figures include a plethora of code and decoder combinations for the \ac{LDPC} code: two flavors of the \ac{DVB-S2} \ac{LDPC} code, one with the outer \ac{BCH} code, and one with just the ``inner'' \ac{LDPC} code. Fig.~\ref{fig:long_medium_rate} shows a \ac{LBP} decoder with $8$ iterations for both setups and a $32$ iteration baseline for the plain \ac{LDPC} code (the outer code has a negligible effect on error correction performance for the higher iteration count). 
Also shown is a density evolution-optimized polar code under plain \ac{SC} decoding and a $(4, 8)$ regular \ac{SC-LDPC} code with coupling width $w=3$ and block length $n=6400$, decoded with window size $D=8$ and one BP iteration per window step. 
Fig.~\ref{fig:long_high_rate} shows similar results for the high-rate scenario. 

In addition to the \ac{FER} shown in all three figures, Fig.~\ref{fig:long_medium_rate} shows the \ac{BER}. While the \ac{FER} is normally the metric to focus on---as a frame error would trigger a repeat request in most scenarios---looking at the \ac{BER} allows to extract some interesting information about convergence behavior.

\section{Candidate I: Polar Codes}\label{sec:polar_candidate}
Polar codes, introduced by Arıkan in \cite{arikan2009}, are the first family of channel codes to provably achieve the capacity of discrete memoryless channels using \ac{SC} decoding with a decoding complexity of $O(N\log N)$. 
Despite their asymptotic optimality, it is remarkable that polar codes excel in the short block length scenario.
This is due to the discovery that pre-transformations, e.g., using \ac{CRC} codes,  effectively eliminate low-weight codewords. 
Then, combined with advanced SC-based decoding schemes, e.g., \ac{SCL} \cite{Dumer2006SCL,talvardy2015} and \ac{AED} \cite{Geiselhart2021AED}, polar codes yield competitive performance in the short block length regime.
For instance, \ac{CRC}-aided polar codes are employed in the control channel in the 5G standard. 
Furthermore, \ac{SC}-based decoders do not suffer from an error floor, enabling polar codes to be used in \ac{HRLLC} scenarios.

\subsection{Code Properties}
A plain polar code is defined by the set of information bits $\mathcal{I}$. Therefore, depending on the chosen decoder and design parameters, a suitable set must be determined.
For plain \ac{SC} decoding, density evolution is an effective algorithm to determine the information bit positions.
All subsequent corresponding code designs have been optimized for a \ac{FER} of $10^{-6}$ using density evolution.

\begin{figure}[t]
	\centering
    {\newcommand\lineWidth{0.75pt}
\newcommand\markSize{2pt}

\begin{tikzpicture}
\begin{axis}[
width=\linewidth,
height=\linewidth,
grid style={dotted,anthrazit},
xmajorgrids,
yminorticks=true,
ymajorgrids,
legend columns=1,
legend pos=north east,   
legend cell align={left},
legend style={fill,fill opacity=0.8},
xlabel={$E_\mathrm{b}/N_0$ in dB},
ylabel={FER},
label style={font=\small},
tick label style={font=\small}, 
legend image post style={mark indices={}},%
ymode=log,
mark size=2.5pt,
xmin=1,
xmax=5.25,
ymin=1e-6,
ymax=1
]

\addplot[color=rot,line width = 0.75pt, solid, mark=o, mark options={solid},mark repeat=1,mark phase=1, mark size = 1.5pt]
table[col sep=comma]{
1.00, 6.758e-01
1.50, 3.973e-01
2.00, 1.766e-01
2.50, 5.969e-02
3.00, 1.533e-02
3.50, 2.935e-03
4.00, 4.084e-04
4.50, 4.084e-05
5.00, 2.977e-06
5.50, 1.613e-07
6.00, 6.080e-09
6.50, 1.504e-10
7.00, 2.278e-12
};
\label{plot:short_SC}
\addlegendentry{\footnotesize Polar SC}; %

\addplot[color=apfelgruen,line width = 0.75pt, solid, mark=o, mark options={solid},mark repeat=1,mark phase=1, mark size = 1.5pt]
table[col sep=comma]{
0.0,	9.568e-01
0.5,	8.253e-01
1.0,	5.496e-01
1.5,	2.966e-01
2.0,    1.180e-01
2.5,	3.247e-02
3.0,	5.697e-03
3.5,	9.123e-04
4.0,	9.090e-05
4.5,	8.019e-06
};
\label{plot:short_LBP}
\addlegendentry{\footnotesize LDPC 5G LBP-8};

\addplot[color=lila,line width = 0.75pt, solid, mark=square, mark options={solid},mark repeat=1,mark phase=1, mark size = 1.5pt]
table[col sep=comma]{
0.00, 7.704e-01
0.50, 6.169e-01
1.00, 3.442e-01
1.50, 1.445e-01
2.00, 4.394e-02
2.50, 9.028e-03
3.00, 1.220e-03
3.50, 1.349e-04
4.00, 9.351e-06
4.50, 6.785e-07
};
\label{plot:short_AED}
\addlegendentry{\footnotesize Polar AE-SC-8}
\addplot[color=mittelblau,line width = 0.75pt, solid, mark=diamond, mark options={solid},mark repeat=1,mark phase=1, mark size = 2.0pt]
table[col sep=comma]{
1.00, 3.665e-01
1.50, 1.333e-01
2.00, 3.174e-02
2.50, 4.929e-03
3.00, 4.434e-04
3.50, 2.452e-05
4.00, 6.988e-07 
}; 
\label{plot:short_SCL}
\addlegendentry{\footnotesize Polar 5G CA-SCL-8}

\addplot[color=black, line width = 0.75pt, solid, mark=none, mark options={solid},postaction={decorate,decoration={raise=1.25pt,text align={left, left indent=2.25cm},text along path,text={|\scriptsize| meta-converse bound}}}, forget plot]
table[col sep=comma]{
0.00, 4.718e-01
0.10, 4.130e-01
0.20, 3.552e-01
0.30, 2.999e-01
0.40, 2.483e-01
0.50, 2.014e-01
0.60, 1.598e-01
0.70, 1.240e-01
0.80, 9.394e-02
0.90, 6.942e-02
1.00, 4.999e-02
1.10, 3.503e-02
1.20, 2.387e-02
1.30, 1.579e-02
1.40, 1.014e-02
1.50, 6.305e-03
1.60, 3.795e-03
1.70, 2.208e-03
1.80, 1.241e-03
1.90, 6.722e-04
2.00, 3.509e-04
2.10, 1.762e-04
2.20, 8.499e-05
2.30, 3.935e-05
2.40, 1.746e-05
2.50, 7.415e-06
2.60, 3.010e-06
2.70, 1.166e-06
2.80, 4.308e-07
2.90, 1.515e-07
3.00, 5.061e-08
3.10, 1.605e-08
3.20, 4.825e-09
3.30, 1.372e-09
3.40, 3.686e-10
3.50, 9.342e-11
3.60, 2.229e-11
3.70, 5.001e-12
3.80, 1.053e-12
3.90, 2.077e-13
}; %

\end{axis}

\end{tikzpicture}
\vspace{-1em}}
	\caption{\footnotesize \ac{FER} vs \ac{SNR} of various \ac{LDPC}- and polar coding schemes of length $N=256$ and rate $R=\nicefrac{1}{2}$.}
    \label{fig:short_codes}
\end{figure}
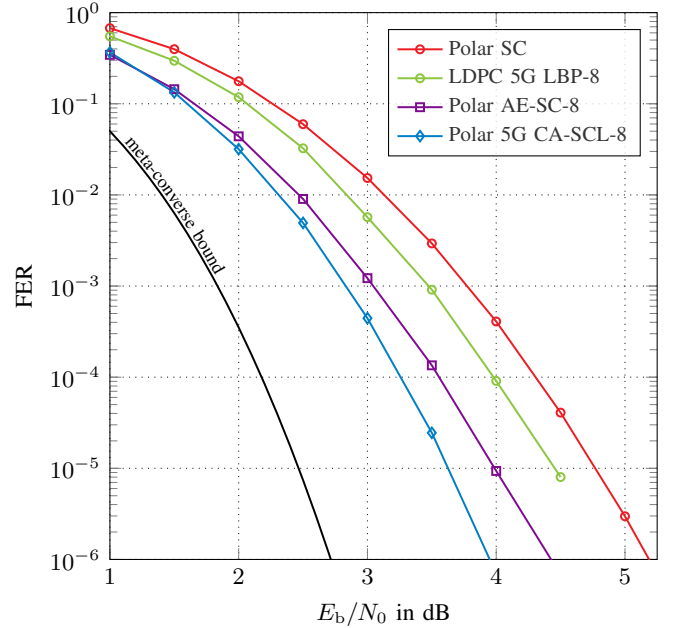

However, there are several other approaches to constructing polar codes with different design choices, e.g., list-aware \cite{polar_code_const_list_decoding}, automorphism-aware design processes, or nested sequence-based code designs \cite{ruebenacke2025NestedPolar}.
Using a nested reliability sequence, as in the 5G case, the description of a polar code is reduced to an ordered list of bit channels.

Polar codes under improved \ac{SC}-based decoding schemes such as \ac{SCL}-decoding and \ac{AED} perform well in short block length scenarios.
Fig.~\ref{fig:short_codes} depicts the \ac{FER} performance of the 5G polar code under \ac{CA} \ac{SCL}-decoding of list size $8$ (\ref{plot:short_SCL}), as well as the \ac{FER} performance of the polar code with $\mathcal{I}_\mathrm{min}=\{31,57\}$ from \cite{geiselhart2021polaraed} under \ac{SC}-based \ac{AED} (\ref{plot:short_AED}). 
Both code designs and decoders show comparative performance and outperform the 5G \ac{LDPC} code with $8$ iterations of \ac{LBP} (\ref{plot:short_LBP}) of the same length, while the plain \ac{SC} decoder (\ref{plot:short_SC}) cannot match the performance of the \ac{LDPC} code. 
Furthermore, for increasing block lengths, polar codes with plain \ac{SC} decoding allow for low-complexity implementation while still providing competitive performance. Further reduction of complexity is possible by using fast-\ac{SSC} decoding.

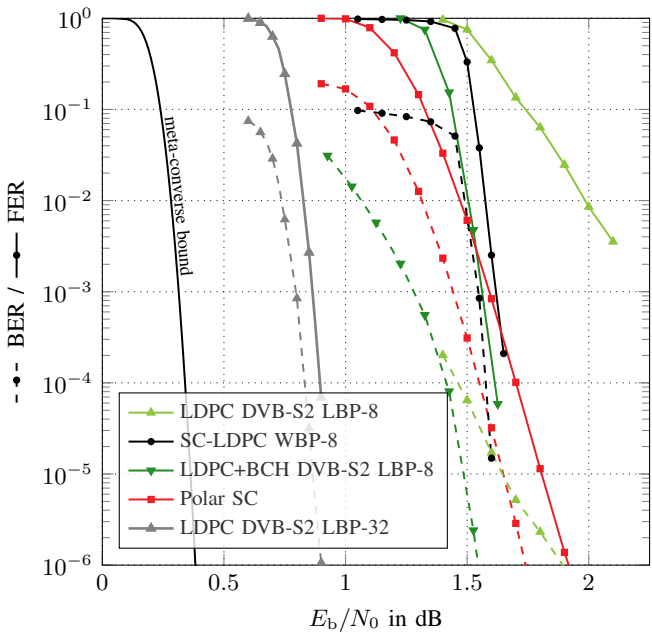
\begin{figure}[htp]
	\centering
    {\newcommand\lineWidth{0.75pt}

\begin{tikzpicture}
\begin{axis}[
width=\linewidth,
height=\linewidth,
grid style={dotted,anthrazit},
xmajorgrids,
yminorticks=true,
ymajorgrids,
legend columns=1,
legend pos=south west,   
legend cell align={left},
legend style={fill,fill opacity=0.8},
xlabel={$E_\mathrm{b}/N_0$ in dB},
ylabel={\ref{plot:BER, N=64k} BER / \ref{plot:BLER, N=64k} FER},
label style={font=\small},
tick label style={font=\small}, 
legend image post style={mark indices={}},%
ymode=log,
mark size=2.5pt,
xmin=0,
xmax=2.25,
ymin=1e-6,
ymax=1
]

\addplot[color=apfelgruen,line width = 0.75pt, solid, mark=triangle*, mark options={solid}, mark size=1.5pt]
table[col sep=comma]{
1.4,    9.627e-01
1.5,    7.531e-01
1.6,    3.459e-01
1.7,    1.346e-01
1.8,    6.345e-02
1.9,    2.465e-02
2.0,    8.547e-03
2.1,    3.547e-03
};
\label{plot:dvbs2_8it}
\addlegendentry{\footnotesize LDPC DVB-S2 LBP-8}

\addplot[color=black,line width = 0.75pt, mark=*, mark options={solid},mark repeat=1,mark phase=1, mark size=1pt]
table[col sep=comma]{
1.05,	0.98096354167
1.15,	0.971953124994
1.25,	0.956640625003
1.35,	0.922786458327
1.45,	0.7763020833237
1.5,	0.3316406249977
1.55,	0.03788411457587
1.6,	0.002515714800108
1.65,	0.000209668803856
};
\label{plot:BLER, N=64k}
\addlegendentry{\footnotesize SC-LDPC WBP-8}

\addplot[color=darkgreen,line width = 0.75pt, mark=triangle*, mark options={solid, rotate=180},mark repeat=1,mark phase=1, mark size=1.5pt]
table[col sep=comma]{
1.2258,	1.000e+00
1.3258,	7.500e-01
1.4258,	1.538e-01
1.5258, 4.757e-03
1.6258, 5.853e-05
};
\label{plot:BLER, N=64k, R=1/2, DVB_BCH}
\addlegendentry{\footnotesize LDPC+BCH DVB-S2 LBP-8}

\addplot[color=rot,line width = 0.75pt, solid, mark=square*, mark options={solid}, mark size=1pt]
table[col sep=comma]{
0.90, 1.000e+00
1.00, 9.864e-01
1.10, 7.884e-01
1.20, 4.186e-01
1.30, 1.453e-01
1.40, 3.309e-02
1.50, 6.066e-03
1.60, 8.370e-04
1.70, 1.013e-04
1.80, 1.143e-05
1.90, 1.381e-06
2.00, 2.183e-07
};
\label{plot:BLER, N=64k, R=1/2, Polar SC}
\addlegendentry{\footnotesize Polar SC}

\addplot[color=mittelgrau,line width = 1pt, mark=triangle*, mark options={solid},mark repeat=2,mark phase=1, mark size=1.5pt]
table[col sep=comma]{
0.6,9.924e-01
0.625,9.556e-01
0.65,8.958e-01
0.675,8.212e-01
0.7,6.316e-01
0.725,4.615e-01
0.75,2.460e-01
0.775,1.043e-01
0.8,4.243e-02
0.825,1.096e-02
0.85,2.694e-03
0.875,4.611e-04
0.9,6.919e-05
};
\addlegendentry{\footnotesize LDPC DVB-S2 LBP-32}
\label{plot:dvbs2_32it}

\addplot[color=mittelgrau, dashed, line width = \lineWidth, mark=triangle*, mark options={solid},mark repeat=2,mark phase=1, mark size=1.5pt]
table[col sep=comma]{
0.6,7.462e-02
0.625,6.741e-02
0.65,5.626e-02
0.675,4.417e-02
0.7,2.884e-02
0.725,1.528e-02
0.75,6.209e-03
0.775,2.343e-03
0.8,8.456e-04
0.825,1.551e-04
0.85,3.183e-05
0.875,4.996e-06
0.9,1.066e-06
};

\addplot[color=rot,line width = \lineWidth, dashdotted, mark=none, mark options={solid}]
table[col sep=comma]{
2.00, 2.274e-07
2.10, 4.678e-08
2.20, 1.190e-08
2.30, 3.481e-09
2.40, 1.124e-09
2.50, 3.915e-10
2.60, 1.458e-10
2.70, 5.773e-11
2.80, 2.406e-11
2.90, 1.043e-11
3.00, 4.632e-12
3.10, 2.080e-12
3.20, 9.351e-13
};

\addplot[color=rot,line width = 0.75pt, dashed, mark=square*, mark options={solid}, mark size=1pt]
table[col sep=comma]{
0.90, 1.915e-01
1.00, 1.678e-01
1.10, 1.081e-01
1.20, 4.624e-02
1.30, 1.261e-02
1.40, 2.334e-03
1.50, 3.102e-04
1.60, 3.213e-05
1.70, 2.869e-06
1.80, 2.081e-07
1.90, 1.350e-08
2.00, 9.074e-10
};

\addplot[color=apfelgruen,line width = 0.75pt, solid, mark=triangle*, mark options={solid}, mark size=1.5pt, dashed]
table[col sep=comma]{
1.4,	2.013e-04
1.5,	6.440e-05
1.6,	1.718e-05
1.7,	5.184e-06
1.8,	2.316e-06
1.9,	9.417e-07
2.0,	2.981e-07
2.1,	1.226e-07
};

\addplot[color=darkgreen,line width = 0.75pt, mark=triangle*, mark options={solid, rotate=180},mark repeat=1,mark phase=1, mark size=1.5pt, dashed]
table[col sep=comma]{
0.9258,	3.106e-02
1.0258,	1.427e-02
1.1258,	5.736e-03
1.2258,	2.029e-03
1.3258,	5.547e-04
1.4258,	8.072e-05
1.5258,	2.408e-06
1.6258,	2.818e-08

};

\addplot[color=black,line width = 0.75pt, solid, mark=*, mark options={solid}, mark size=1pt, dashed]
table[col sep=comma]{
1.05,		0.09734347874689
1.15,		0.090880636945488
1.25,		0.0833029296937712
1.35,		0.0731524414117884
1.45,		0.0509816840248953
1.55,		0.0008476806637386
1.6,		1.49073575144483e-05
}; 
\label{plot:BER, N=64k}

\addplot[color=black,line width = 0.75pt, solid, mark=none, mark options={solid}, postaction={decorate,decoration={raise=1.25pt,text align={left, left indent=1.68cm},text along path,text={|\scriptsize| meta-converse bound}}}, forget plot]
table[col sep=comma]{
0.00, 1.000e+00
0.05, 9.993e-01
0.10, 9.784e-01
0.11, 9.630e-01
0.12, 9.394e-01
0.13, 9.055e-01
0.14, 8.592e-01
0.15, 7.995e-01
0.16, 7.266e-01
0.17, 6.425e-01
0.18, 5.506e-01
0.19, 4.558e-01
0.20, 3.633e-01
0.21, 2.781e-01
0.22, 2.039e-01
0.23, 1.430e-01
0.24, 9.565e-02
0.25, 6.098e-02
0.26, 3.699e-02
0.27, 2.132e-02
0.28, 1.167e-02
0.29, 6.060e-03
0.30, 2.982e-03
0.31, 1.390e-03
0.32, 6.130e-04
0.33, 2.558e-04
0.34, 1.009e-04
0.35, 3.760e-05
0.36, 1.324e-05
0.37, 4.400e-06
0.38, 1.380e-06
0.39, 4.083e-07
0.40, 1.139e-07
0.45, 8.186e-11
0.50, 1.256e-14
};

\end{axis}

\end{tikzpicture}
\vspace{-1em}}
	\caption{\footnotesize \ac{FER} and \ac{BER} vs \ac{SNR} of various \ac{LDPC}- and polar coding schemes of length $N=65536$ and rate $R=\nicefrac{1}{2}$.}
    \label{fig:long_medium_rate}
\end{figure}

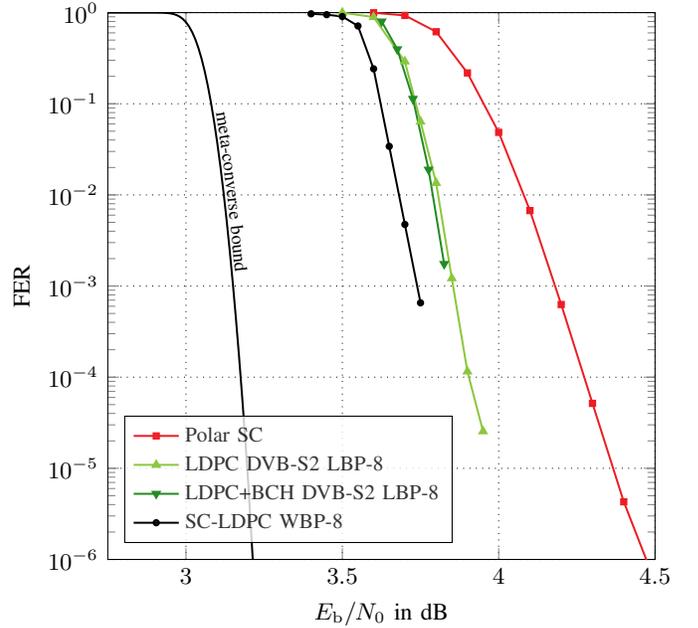
\begin{figure}[htp]
	\centering
    {\newcommand\lineWidth{0.75pt}

\begin{tikzpicture}
\begin{axis}[
width=\linewidth,
height=\linewidth,
grid style={dotted,anthrazit},
xmajorgrids,
yminorticks=true,
ymajorgrids,
legend columns=1,
legend pos=south west,   
legend cell align={left},
legend style={fill,fill opacity=0.8},
xlabel={$E_\mathrm{b}/N_0$ in dB},
ylabel={FER},
label style={font=\small},
tick label style={font=\small}, 
legend image post style={mark indices={}},%
ymode=log,
mark size=2.5pt,
xmin=2.75,
xmax=4.5,
ymin=1e-6,
ymax=1
]

\addplot[color=rot,line width = \lineWidth, solid, mark=none, mark=square*, mark options={solid}, mark size=1pt]
table[col sep=comma]{
3.60, 9.965e-01
3.70, 9.319e-01
3.80, 6.172e-01
3.90, 2.180e-01
4.00, 4.857e-02
4.10, 6.733e-03
4.20, 6.267e-04
4.30, 5.163e-05
4.40, 4.293e-06
4.50, 5.660e-07
4.60, 9.712e-08
};
\addlegendentry{\footnotesize Polar SC};

\addplot[color=apfelgruen,line width = \lineWidth, mark=triangle*, mark options={solid}, mark size=1.5pt]
table[col sep=comma]{
3.5,	1.000e+00
3.6,	8.929e-01
3.7,	2.911e-01
3.75,	6.371e-02
3.8,	1.352e-02
3.85,	1.216e-03
3.9,    1.155e-04 
3.95,   2.546e-05 
};
\addlegendentry{\footnotesize LDPC DVB-S2 LBP-8}

\addplot[color=darkgreen,line width = \lineWidth, mark=triangle*, mark options={solid, rotate=180}, mark size=1.5pt]
table[col sep=comma]{
3.6258,	8.000e-01
3.6758,	3.964e-01
3.7258,	1.133e-01
3.7758, 1.894e-02
3.8256, 1.747e-03
};
\addlegendentry{\footnotesize LDPC+BCH DVB-S2 LBP-8};

\addplot[color=black,line width = \lineWidth, mark=*, mark options={solid},mark repeat=1,mark phase=1, mark size=1pt]
table[col sep=comma]{
3.4,	0.97065104165
3.45,	0.94966145833
3.5,	0.90677083331
3.55,	0.7139583333467
3.6,	0.2417708333137
3.65,	0.03408854166298
3.7,	0.0047415865372
3.75,	0.0006542968744
};
\addlegendentry{\footnotesize SC-LDPC WBP-8};

\addplot[color=black,line width = \lineWidth, solid, mark=none, mark options={solid}, postaction={decorate,decoration={raise=1.25pt,text align={left, left indent=3.3cm},text along path,text={|\scriptsize| meta-converse bound}}}, forget plot]
table[col sep=comma]{
2.50, 1.000e+00
2.85, 1.000e+00
2.90, 9.996e-01
2.91, 9.990e-01
2.92, 9.977e-01
2.93, 9.950e-01
2.94, 9.899e-01
2.95, 9.806e-01
2.96, 9.649e-01
2.97, 9.401e-01
2.98, 9.032e-01
2.99, 8.519e-01
3.00, 7.850e-01
3.01, 7.031e-01
3.02, 6.093e-01
3.03, 5.085e-01
3.04, 4.068e-01
3.05, 3.109e-01
3.06, 2.264e-01
3.07, 1.566e-01
3.08, 1.027e-01
3.09, 6.369e-02
3.10, 3.733e-02
3.11, 2.064e-02
3.12, 1.075e-02
3.13, 5.275e-03
3.14, 2.433e-03
3.15, 1.055e-03
3.16, 4.294e-04
3.17, 1.641e-04
3.18, 5.882e-05
3.19, 1.977e-05
3.20, 6.229e-06
3.25, 7.275e-09
3.30, 1.594e-12
3.35, 6.405e-17
};

\end{axis}

\end{tikzpicture}
\vspace{-2em}}
	\caption{\footnotesize \ac{FER} vs \ac{SNR} of various \ac{LDPC}- and polar coding schemes of length $N=65536$ and rate $R=\nicefrac{8}{9}$.}
    \label{fig:long_high_rate}
\end{figure}

Despite these advantages, polar codes have limitations: 
In contrast to iterative decoding schemes, \ac{SC}-based decoding schemes do not inherently allow for an on-the-fly adaption of the complexity-performance trade-off. 
Furthermore, hardware design and practical implementations for long block lengths are not yet widely studied.
In addition, adapting polar codes to block lengths $N\neq2^n$ can be more involved.
While several solutions have been proposed to address length matching, such as \cite{Niu2013PolarRateCompatible,Wang2014PolarRateCompatible,Bioglio2017PolarRateCompatible,Saber2015PolarRateCompatible}, an efficient and elegant length-matching approach has yet to be found. 
For longer finite block lengths, as depicted in Fig.~\ref{fig:long_medium_rate} and Fig.~\ref{fig:long_high_rate}, plain \ac{SC} decoding (\ref{plot:BLER, N=64k, R=1/2, Polar SC}) fails to achieve a steep error-rate curve. 
Minimizing the gap to capacity through improved decoding algorithms and code constructions is essential for polar codes to remain competitive among state-of-the-art error-correcting codes for large finite block lengths and higher decoding complexity. Contrary to popular belief, recent results suggest that plain polar codes demonstrate competitive performance in a decoding complexity-constrained environment \cite{krieg2025long}. However, they lack the ability to improve performance in non-complexity-constrained setups, as the decoder complexity of \ac{SC} is constant for given code parameters and can not be adjusted.

\subsection{Implications for Hardware Implementation}

Polar codes initially appear attractive due to their inherent structural regularity, high locality, and node-level parallelism within the nodes of the 
\ac{PFT} \cite[Fig. 1]{simplifiedSC}. These properties suggest the potential for efficient hardware realization. However, the recursive nature of the polarization transformation results in a sequential decoding process. Specifically, decoding polar codes can be viewed as a depth-first traversal through the \ac{PFT}, inherently imposing a serial decoding sequence. This sequential behavior poses significant challenges for achieving architectures with high throughput and low latency.  A second challenge arises from the uneven distribution of operations across different layers of the \ac{PFT}. Specifically, the number of node operations decreases by half with each subsequent layer in the tree, starting from the top. Consequently, the hardware resources allocated for peak processing capacity at the root level remain partially underutilized in deeper tree layers. This inherent imbalance in computational workload across the tree layers complicates achieving consistently high hardware utilization throughout the decoding process. Only fully unrolled and pipelined architectures, which directly map all nodes and their corresponding operations onto dedicated hardware, effectively mitigate these challenges and can achieve very high throughput. However, fully unrolled, pipelined architectures have very limited flexibility and scalability, making it difficult to efficiently support variations such as different block lengths. A third challenge arises in advanced polar decoding schemes such as SCL, where e.g. sorting operations and dynamic list-length management are essential. These requirements introduce additional control flow into the architecture, increasing hardware complexity, leading to reduced energy efficiency, and resulting in larger chip areas.

\subsection{Promising Technologies}
Advances in polar code research have led to new approaches that improve performance for short and long code lengths and increase the feasibility of a unified polar coding scheme. This subsection highlights key techniques that address the challenges of polar codes and provides insight into novel approaches to polar coding.
\subsubsection{Pre-Transformation}
While polar codes are capacity-achieving for infinite length \cite{arikan2009}, they are known to have poor weight enumerators in the finite length regime \cite{less1o5wmin}\cite{less2wmin}.
As near-\ac{ML} decoding techniques are available, e.g., \ac{SCL} or \ac{AED}, the weight enumerators are directly connected to the decoding performance.
Thus, the performance can be greatly improved by using a pre-transform that expurgates low- or minimum-weight codewords.
Common types of pre-transforms are \ac{CRC}s \cite{talvardy2015}, convolutional codes \cite{arikan2019}, and repetition codes (typically referred to as row-merging) \cite{Gelincik2022}\cite{rowmerged}. 

\subsubsection{List Decoding} 
A common near-\ac{ML} decoding technique is \ac{SCL} decoding \cite{talvardy2015}, where at each decision in the SC decoder, both paths are followed.
If the number of paths exceeds the list size $L$, only the $L$ best paths according to a path metric are kept.
This can be viewed as using multiple SC decoders in parallel that collaborate to find the best result.
However, this collaboration comes with control flow overhead for path management and sorting.
On the bright side, \ac{SCL} decoding harmonizes well with a possible pre-transformation.
The constraints of the pre-transform can be considered by interpreting them as dynamic frozen bits. 
Moreover, the main factors in code design are \ac{ML} performance and sub-channel polarization.

\subsubsection{Ensemble Decoding}
\ac{AED} for polar codes \cite{Geiselhart2021AED} is a technique where multiple SC decoding attempts are processed in parallel and independently. 
Subsequently, the best attempt is chosen.
Especially in the short block length regime, this leads to near-\ac{ML} performance.
The advantage, compared to \ac{SCL}, is that the constituent decoders work independently, which avoids control flow overhead. Thus, the small performance gain of \ac{SCL} shown in Fig.~\ref{fig:short_codes} comes at a high computational cost, further demonstrating the capabilities of ensemble-based decoders in the short block length regime. 
The diversity of the ensemble, meaning the possibility of different outcomes per decoder, is enabled by the symmetries of the code.
While polar codes tend to have such symmetries, they are not guaranteed and the code design must explicitly consider them.
Furthermore, the code must have good \ac{ML} performance and good SC performance.
However, it turns out that good polarization and symmetry are conflicting goals, especially for larger block lengths.
    
\subsubsection{Simplified Decoding Algorithms}
Decoding efficiency is key. One way to reduce the complexity in \ac{SC} decoding is the simplification of the decoding algorithm by making use of the code structure.
Whenever a subcode with a specific structure is to be decoded, e.g., rate-1, single parity-check, or repetition, \ac{SSC} employs a specialized, low-complexity decoder \cite{simplifiedSC}\cite{ssc_rep_spc}.
The more of these special subcodes are present in the code design, the more efficient the decoding.

\subsubsection{Rate-Flexible Long Codes}
For long block lengths, the gain of \ac{AED} compared to \ac{SC} diminishes, due to conflicts in code design, and \ac{SCL} decoding becomes too complex. 
Thus, one has to resort to plain \ac{SC} decoding.
The design criterion for \ac{SC} decoding is the polarization of the sub-channels, which can be calculated with density evolution \cite{polarde}.
The order in which the sub-channels reach the target error rate as the \ac{SNR} increases directly creates a rate-flexible reliability sequence for a given length.

\subsubsection{Nesting}
Nesting allows for length flexibility of polar codes using a single reliability sequence \cite{nestedpolar}.
Given a reliability sequence for a longer length, the reliability sequence for shorter lengths can be extracted by considering only the sub-channels indices within the shorter block length.

\subsubsection{Hybrid ARQ}
For energy and spectral efficient transmission of large payload sizes, a code must support incremental redundancy \ac{HARQ}. For polar codes, \ac{HARQ} can be implemented by repeating some information symbols in a larger polar code for subsequent transmissions \cite{ma2017polarharq} \cite{yuan2018polarharq}.
The resulting code is similar to a row-merged polar code \cite{rowmerged} and can be decoded alike.

\subsection{Potential Code Design}

A code design for 6G based on polar codes should contain or enable the mentioned technologies. 
The goal is to have a nested reliability sequence that follows density evolution for \ac{SC} decoding for long block lengths, where polarization is sufficiently good, while the contained sequence for short block lengths (e.g., $N \le 512$ bits) where the code's \ac{ML} performance plays a more important role, is optimized for \ac{AED} or \ac{SCL}.
It turns out that a good code for \ac{AED} is also a good code for \ac{SCL}, as codes with a rich automorphism group typically have good distance properties.
Thus, in light of the need to enable different decoders, i.e., \ac{AED} and \ac{SCL}, a code design should consider \ac{AED}.
Recent work has shown that both rate- and length-flexible polar codes with symmetries can be constructed \cite{geiselhart2023ratecompatible, ruebenacke2025NestedPolar}, compatible both with \ac{AED} and \ac{SCL} decoding.

From a protocol level, an outer code for error detection is required. This outer code can be designed as a suitable pre-transformation for both \ac{AED} and \ac{SCL}-based decoding, e.g., in the form of row-merging \cite{rowmerged} rather than a \ac{CRC}---the outer code used in 5G.

\section{Candidate II: LPDC Codes}\label{sec:LDPC_candidate}
\ac{LDPC} codes together with \ac{BP} decoding are highly effective in terms of error correction, especially for large block lengths. \ac{LDPC} codes, together with BP decoding, are known for their low latency and hardware simplicity, making them a popular choice in various communication systems, including 5G. However, for shorter block lengths, or when limited to a small number of iterations, \ac{LDPC} codes still show room for improvement.

\subsection{Code Properties}
A \ac{LDPC} code is characterized by its sparse \ac{PCM} $\boldsymbol{H}$.
The design of $\boldsymbol{H}$ critically affects performance, with density evolution, \ac{EXIT} charts, and protograph-based constructions serving as fundamental methods to optimize code parameters for iterative \ac{BP} decoding.  

Recent advances include spatially-coupled \cite{kudekar2011SC} and \ac{QC}-\ac{LDPC} designs \cite{fossorier2004quasicyclic}, which enhance error correction capabilities while maintaining hardware-friendly structures. For example, 5G adopts \ac{QC}-\ac{LDPC} codes for their modularity and compatibility with layered decoding.

\ac{LDPC} codes excel in long block-length regimes, where \ac{BP} decoding with a sufficient number of iterations achieves near-capacity performance with linear-time complexity. \ac{BP} decoding dominates practical implementations because its iterative nature allows flexible trade-offs between latency and accuracy, enabling dynamic adaptation to channel conditions---a clear advantage over non-iterative decoders. However, without sufficient iterations \ac{BP} falls short of its potential, as can be seen in Fig.~\ref{fig:long_medium_rate}: The additional $24$ iterations provide a significant performance gain for the \ac{DVB-S2} \ac{LDPC} code (\ref{plot:dvbs2_8it} vs. \ref{plot:dvbs2_32it}). Fig.~\ref{fig:long_medium_rate} visualizes an interesting fact about the low-iteration decoder behavior of \ac{LDPC} codes: the \ac{DVB-S2} \ac{LDPC} code (without its outer \ac{BCH} code) exhibits a significant \ac{FER}-performance gap relative to the polar code (\ref{plot:BLER, N=64k, R=1/2, Polar SC}), yet achieves competitive \ac{BER} performance. This suggests that the \ac{BP} decoder converges quickly on most bit positions but requires additional iterations to resolve residual errors---which is a computationally expensive process. 
To maintain a satisfactory performance in low-iteration regimes, pairing \ac{LDPC} codes with an outer code (\ref{plot:BLER, N=64k, R=1/2, DVB_BCH}) offers a pragmatic solution that preserves higher-iteration \ac{FER} performance with less computational overhead.

Another approach for promising performance at low complexity is \ac{SC-LDPC} coding (\ref{plot:BLER, N=64k}); also shown in Fig.~\ref{fig:long_medium_rate} and Fig.~\ref{fig:long_high_rate}. 

Despite their strengths, \ac{LDPC} codes face challenges: Error floors at high \ac{SNR} regimes persist due to trapping sets \cite{vasic2009trappingsets}, limiting their use in high reliability applications. Achieving competitive error rates with \ac{LDPC} codes at short block lengths remains an open challenge, as displayed in Fig.~\ref{fig:short_codes}. The performance of \ac{LDPC} codes in this short block-length regime often lags behind alternatives such as polar codes, requiring custom designs (which often conflict with flexibility requirements) or auxiliary techniques. For example, a novel contribution \cite{shen2025universalBP} finds \acp{PCM} better suited for \ac{BP}, thus allowing \ac{BP} to be used as a decoder that can perform close to \ac{ML} for arbitrary block codes. 

\subsection{Implications for Hardware Implementation}

\ac{LDPC} codes are typically encoded based on their \ac{PCM}.
Particularly, structured \ac{LDPC} codes allow for systematic encoding, and the encoder implementation is straightforward due to lower-triangular matrices. %

Architectures for iterative \ac{BP} decoding are based on approximations of the \ac{SPA}, such as the \ac{NMS} or \ac{OMS} algorithms. 
These algorithms replace the multiplications of the SPA by using less complex minimum-searches followed by a normalization factor or an offset, respectively. 
This approximation reduces the implementation complexity but slightly impacts the error correction performance.

\ac{BP} decoding is based on an iterative exchange of messages between variable and check nodes in the Tanner graph of the code. 
The inherent parallelism on the node level makes LDPC codes well-suited for high-throughput requirements. 
Thus, decoder architectures are widely applied using either partial or full parallelism. 
On the one hand, partial-parallel decoder architectures allow for flexible reconfiguration for different codeword lengths and code rates.
This reconfiguration requires to be considered at design time in the controller configuration, the memory size and the flexibility of the interleaver network for message exchange. 
On the other hand, full-parallel decoding architectures achieve throughputs up to 100 Gbit/s and reduced latency for a high SNR, which is achieved at the cost of reduced flexibility. 
However, full-parallel decoding architectures are only feasible for a medium codeword length ($<$ 2000 bit). 
To achieve a throughput far beyond 100 Gbit/s, unrolling and pipelining of the iterations can be applied but the number of iterations is limited ($<$ 20). 

For both types of decoder architectures, partial-parallel and full-parallel, the irregularity and limited locality of the LDPC code, reflected in the Tanner graph edges, cause challenges. 
In particular, these challenges are memory access conflicts for partial-parallel decoder architectures and routing congestions for full-parallel decoder architectures
potentially increasing decoding latency, reducing throughput, and increasing energy consumption. 
A layered decoding schedule can increase regularity and locality in a decoder architecture, particularly for CN regular LDPC codes. 
For layered decoding, the processing of the Tanner graph is serialized and the nodes partially updated, resulting in faster convergence of the decoding.
Due to the lower number of iterations, the area efficiency and energy efficiency are improved at the cost of an increased decoding latency.

A challenge of \ac{LDPC} code decoders is their non-deterministic latency due to the required number of iterations. 
Thus, an upper limit must be defined, particularly for unrolled and pipelined decoder architectures which inherently limits the error correction performance in the waterfall region. 
Another challenge is high-throughput decoding of long codeword length. 
So far, only pipelined sliding window decoders for \ac{SC-LDPC} codes seem to be a feasible solution as they provide locality due to their band diagonal matrix structure and thus allow for a small decoding window size. 
As the pipelined sliding window decoder architecture does not perform iterations within the decoding window, traditional iteration control techniques to reduce the power consumption can not be applied.
A possible solution to reduce the power consumption depending on the \ac{SNR} is the dynamic adaption of the window size. 

\subsection{Promising Technologies}

\subsubsection{Ensemble Decoding for Short block lengths}
Since \ac{BP} decoders behave sub-optimally, different decoders can cooperate to produce a collective estimate. Similar to the ensemble approach for polar codes, each \ac{BP} algorithm in the ensemble decodes independently. Subsequently, they can be implemented in parallel. This parallel implementation, together with the better error performance at a lower number of iterations (per ensemble member), enables a lower decoding delay and is thus beneficial for low-latency use cases. 

For instance, ensembles of \ac{BP} decoders can be built from automorphisms, from different schedules of the layered decoder, by adding noise \cite{krieg2024ed} or by adding linearly independent rows to the \ac{PCM} to constitute subcode decodings \cite {mandelbaum2025sced}. Ensembles built from automorphisms require special code design, which typically significantly degrades standalone performance compared to \ac{LDPC} codes built without this constraint. Scheduling ensembles and subcode ensembles do not impose such constraints on code design and are typically applicable to all \ac{LDPC} codes. Hereby, scheduling ensembles require a layered \ac{BP} decoder. Whether it is beneficial to fine-tune the code design for better ensemble performance or simply to use a scheduling ensemble is still an open question.
 
\subsubsection{Rate-Adaptivity}
Generally, rate-adaptivity of any code can be achieved via puncturing and shortening. A \ac{PBRL} code design is advantageous in terms of rate flexibility, as the rate can be adjusted by utilizing the extension part of the \ac{PCM}. This approach enables spectrally efficient \ac{HARQ} systems. 

An alternative approach involves using a distribution matcher along with a fixed rate and block length protograph-based \ac{LDPC} code. For instance, rate-adaptive MacKay-Neal codes may be considered for very high-throughput connections, as discussed in \cite{Zahr2025}.

\subsubsection{Hardware friendly LDPC code design}
\ac{LDPC} codes offer significant flexibility in their design and application. Various design methods for \ac{LDPC} codes focus primarily on enhancing error-rate performance; however, these methods often conflict with the hardware requirements outlined in Section \ref{sec:iv.b.implementation_perspective}. Consequently, it is crucial to consider hardware implementation during the code design process, transforming it into a multi-objective optimization problem.

To minimize power consumption, the algorithmic complexity of decoding must be reduced. The complexity of a \ac{BP} decoder is related to the number of decoding iterations and the number of edges in the Tanner graph. Therefore, achieving low complexity necessitates a lower average check node degree, which serves as a proxy for the number of edges. Codes that are Pareto optimal in terms of performance and decoding complexity can be developed, as demonstrated in \cite{finite_exit_chart_koike}.

A structured \ac{PCM} can facilitate a local and scalable hardware implementation. For example, the construction of a quasi-cyclic matrix enables partially parallel and layered implementations due to independent edge connections of some check nodes.  Considering layered implementations, reducing the overlap and congestion of the decoding paths reduces the idle time of the check node processing units caused by memory access. To this end, the design of a \ac{PCM} can be constrained on a low co-dependency of subsequently scheduled check nodes \cite{design_of_5g_richie}. Furthermore, codes optimized solely for maximum performance often result in highly irregular code distributions, which can decrease hardware efficiency and hinder scalability. Thus, it is essential to limit the degree of irregularity among node degrees during the distribution optimization process. Lastly, for high throughput implementations the maximum number of iterations is limited since unrolled, fully pipelined architectures are used. Thus, for a hardware friendly code design, the performance for finite iterations has to be taken into account.

Another key factor to enable scalability is locality in the decoding process. Local computations limit data transfer and are therefore advantageous from a hardware perspective. For this reason, techniques such as spatial coupling or global coupling support scalability. Prioritizing local edge connections during the \ac{PCM} generation process can further enhance this locality. 

The considerations of imposing structure on the \ac{PCM} also aim towards achieving regularity which enables the reuse of building blocks in hardware.

The practicability of the coding scheme is mainly determined by its simplicity. Codes that allow for easy adjustment of both code rate and block length are desirable to simplify hardware implementation. For instance, \ac{PBRL} codes meet this requirement. The block length is adjusted by modifying the size of the circulant matrices used in quasi-cyclic construction. The rate matching via puncturing of check nodes enables \ac{HARQ} without massive control flow. 

\subsubsection{SC-LDPC Codes}
\Ac{SC-LDPC} codes exhibit improved error correction performance over conventional \ac{LDPC} block codes, achieving results comparable to the \ac{DVB-S2} \ac{LDPC} code in Fig.~\ref{fig:long_medium_rate} at similar complexity, while outperforming it in Fig.~\ref{fig:long_high_rate}---without the need for an outer code. This highlights their potential to reduce residual errors through spatial coupling, thereby reducing reliance on auxiliary components such as \ac{BCH} codes. This improvement has been both theoretically predicted in the asymptotic regime \cite{lentmaier_2010_SC_threshold, Kudekar_2011_SC_TS, Kudekar_2013_SC_CA} and demonstrated in practice for very large codeword lengths \cite{chapter7_mukherjee_2020}. \ac{SC-LDPC} codes are constructed using a series of independent, usually identical \ac{LDPC} block codes, where edges in each graph are connected to a specified number of neighboring nodes---a process known as edge spreading. Existing design tools developed for \ac{LDPC} block codes, such as density evolution and \ac{EXIT} charts, remain applicable to \ac{SC-LDPC} codes. Edge spreading can be performed after lifting the base graph or the base graph level. In the former case, the resulting \ac{SC-LDPC} code design is less constrained, which enables potentially better error-rate performance, while in the latter, it preserves its quasi-cyclic structure and maintain \ac{HARQ} capabilities, e.g., if the underlying code is raptor-like as in \cite{ren2024esrl}. 

After extensive research, a near-optimal decoder for \ac{SC-LDPC} codes---known as the windowed decoder---has been widely adopted \cite{lentmaier_2011_windowdecoder, Iyengar_2012_windowed_decoder}. The core of the windowed decoder is a \ac{BP} decoder, which operates by sliding across the spatially-coupled chain of \ac{LDPC} codes, outputting data with a delay proportional to the window size. The error-correction capability of the decoder is closely tied to the product of the window size and the codeword length of the underlying \ac{LDPC} codes \cite{chapter7_mukherjee_2020}. Thus, the window size offers a trade-off between decoding complexity and performance.

\subsection{Potential Code Design}
A code design for 6G based on \ac{LDPC} codes should contain or enable the mentioned technologies. The goal is to have a code design that meets the stringent hardware constraints while providing close-to-capacity error-rate performance across a variety of block lengths and code rates. For short block lengths, the code design should be focused on supporting special decoder architectures, such as ensemble decoders. For increasing block lengths, the scalability and hardware constraints become increasingly predominant, and thus, the code design should focus on structure and locality. A possible design might morph from block codes, over \ac{PBRL} codes, to spatially-coupled \ac{PBRL} codes, with increasing block length. For very short block lengths, \ac{ML} decoders are feasible and with modifications to the \ac{PCM}, \ac{BP} serves as a universal decoder that performs close to \ac{ML} \cite{shen2025universalBP}. Therefore, the code design in the short block length regime should provide good \ac{ML} performance and follow a simple or unified construction to enable flexibility. For increasing block length, the code design should shift focus towards complexity reduction while sticking to a unified construction.

\section{Summary and our Vision}\label{sec:conclusion}
In this work, we compared polar and \ac{LDPC} codes in terms of theoretical foundations, decoding performance, and hardware implications and identified key technologies for error-correcting capability, flexibility, and implementation complexity. 
We found that hardware-aware code design is critical to realizing the full potential of a unified coding scheme that combines a simple description with satisfactory performance, while keeping design, description, and hardware overhead to a minimum. 
Critical to this unification is a holistic design methodology in which code construction and decoder architecture evolve jointly to trade the conflicting demands of performance, adaptability, and efficiency.

To address this interplay between code design and hardware efficiency, we envision a modular multi-core architecture that dynamically adapts to varying block lengths. 
For short block lengths, multiple decoder cores operate independently within an ensemble or list decoding framework, maximizing hardware utilization through parallelism. As block lengths increase, cores can be merged via reconfigurable interconnections, balancing the number of constituent decoders against the resource allocation per decoder. For the longest code lengths, all cores can be combined into a single decoder because ensemble and list decoding approaches provide negligible gains for codes of this length. This flexibility ensures efficient resource allocation across multiple encoding scenarios. 

For such a modular design, ensemble decoders--and their inherent independence of the decoding paths--are better suited than traditional list decoders, which rely heavily on control flow due to their cooperative behavior. 
While both \ac{SCL} and \ac{AED} achieve competitive error rates for polar codes, ensemble decoders naturally align with the constituent core paradigm: each decoder operates independently, minimizing control overhead between cores. This independence simplifies scalability and efficiency, especially in short block-length regimes where parallelism is one of the main drivers for competitive decoder implementation \cite{aed23kestel}. In contrast to polar codes, \ac{LDPC} codes do not come with a built-in list decoder but with various flavors of ensembles (e.g., automorphism-based or schedule-based) \cite{krieg2024ed}. 

In addition, the integration of dynamic core activation mechanisms---such as round-robin scheduling---with the modular architecture could improve thermal management during partial hardware utilization \cite{scholl_saturated_2016}. By distributing workloads over the cores, hot spots are mitigated, and heat is better distributed across the chip area.
Both \ac{LDPC} and polar codes allow for this architectural vision. 

For polar codes, the decoder could consist of a single \ac{SSC} decoder for long block lengths and an ensemble of \ac{SSC} decoders using automorphisms of the pre-transformed code in the short block length regime. 

For \ac{LDPC} codes, on the other hand, long block lengths could consist of windowed \ac{BP} decoding, while short blocks could, for example, use ensembles of diversely scheduled \ac{BP} decoders or of subcode decodings. 
Recent work, such as \cite{shen2025universalBP}, further improves short block-length \ac{BP}-decoding by optimizing the graph structure itself---removing cycles and small trapping sets---to improve \ac{BP} convergence. A universal short-block decoding architecture would unify decoding across code families and is in line with our goal of a hardware-friendly, competitive, unified coding scheme.

\bibliographystyle{IEEEtran}
\bibliography{IEEEabrv,bib}

\end{document}